# ABSORPTION AND DESORPTION CURRENTS IN POLYSTYRENE DOPED WITH DR1

S. N. Fedosov[1], T. A. Revenyuk[1], J. A. Giacometti[2]

[1]Odessa National Academy of Food Technologies, Odessa, Ukraine
[2]Sao Paulo State University, Presidente Prudente, Brazil

## Abstract

Isothermal polarization and depolarization currents in polystyrene doped with DR1 chromophore molecules have been measured and analyzed in order to study the electrical relaxation of the poled order in this *guest-host* system. It has been shown that the superposition principle is applicable to the polymer system under study at temperatures lower than the glass transition temperature indicating that the doped polystyrene can be considered electrically as a linear system with a negligible electric conductivity. It has been found that the response function is not the exponential one and characterized by a broad distribution of relaxation times. By comparing isothermal absorption and desorption currents, the practical information was obtained on the best poling conditions for getting high and stable residual polarization.

## 1. Introduction

According to scientific predictions, organic polymers will gradually replace their inorganic counterparts in optoelectronic devices with non-linear optical (NLO) polymers being one of the most promising new materials [1-4]. The system must not possess a center of symmetry, in order to have the NLO properties. Therefore, in guest-host materials, the NLO chromophores dissolved in a glassy polymer matrix must be permanently aligned for breaking the symmetry and producing the residual polarization. This is most easily accomplished by applying a DC electric field to the polymer at the elevated temperature at which the chromophore dipoles can be readily oriented. The thin polymer film are usually placed between two parallel electrodes with the polar axis perpendicular to the film planes (sandwich poling) [5].

Performance of the poled material is limited by the structural relaxation inherent to all glassy polymers leading to eventual randomization of the preferential orientation of the chromophore molecules [6-8]. The constraints on the degree to which the poling-induced orientation of chromophores can be maintained over long periods of time above room temperature are not well understood. It is also not clear what theoretical description is appropriate for the polymer relaxation below the glass transition temperature ($T_g$). One does not understand what poling and physical aging protocols and methods are most appropriate for maximum microstrustural alignment and increased temporal stability, because the phenomena of formation and relaxation of the poled order in NLO polymers are far from being comprehensively studied and understood.

The dipole orientational relaxation is often probed by the second-harmonic generation (SHG) [2,8]. Results obtained by this method are normally do not fit to a simple exponential Debye model corresponded to one relaxation time, therefore either the Kohlrausch-Williams-Watts (KWW) stretched exponential function, or the bi-exponential function is used, both characterizing a broad distribution of relaxation times.

At the same time, we believe that the relaxation processes in NLO polymers can be studied by purely electrical methods adopted from the well developed field of the electret research [5]. As an example, in this paper we report on application of isothermal polarization and depolarization current measurements to investigate formation and relaxation of the poled order in amorphous polystyrene (PS) doped with the Disperse Red (DR1) chromophore molecules. We consider this system as a model one on which some features of the relaxation common for all guest-host NLO polymers can be obtained. Electrical relaxation and related phenomena have been previously studied only in pure PS, but even these data are not comprehensive [9-10].

## 2. Experimental procedures

Atactic amorphous PS ($M_w \cong 250\ 000$) was obtained by purification of a commercial resin. Samples were prepared from a mixture of PS and the Disperse Red 1 (DR1) dissolved in chloroform. The solution was spread over a glass plate and after drying the film was treated at 100 °C for 24 h under vacuum. Then the film was removed by immersing the plate into water. The content of the dye was controlled from 0.5% to 2.5% in the films whose thickness was about 20 μm.

The films were characterized using X-ray diffraction and differential scanning calorimetry (DSC). It has been found that the polymer films were entirely amorphous, with a glass transition temperature $T_g$ around 90 °C. Aluminum electrodes of 2 cm² were deposited on both surfaces of the sample by thermal evaporation in vacuum. Prior to performing polarization and depolarization measurements all samples were short circuited and slowly heated to 120 °C (above $T_g$) in order to remove stray polarization and space charge.

During the isothermal experiments, samples were kept at the desired polarization temperature $T_p$ for at least 15 min and then the poling voltage $V_p$ was applied for a definite time $t_p$ while the poling current $I_p$ was continuously recorded. Immediately after completion of poling, the sample was short circuited and the depolarization current $I_d$ was measured and recorded accordingly. Although $I_p$ and $I_d$ flow in opposite directions, they are both presented as positive currents hereafter in order to facilitate comparison of the currents.

## 3. Results and discussion

### 3.1. Isothermal polarization and depolarization currents

In the absence of conductivity, the isothermal polarization current flowing through a dielectric after application of the constant voltage, called also the absorption current, is a pure displacement one depended on the applied poling field $E$ is assumed to be constant and the so called response function $\boldsymbol{j}(t)$ characterizing the definite dielectric at a constant temperature. Thus, the polarization current density can be presented as

$$j_p(t) = E \cdot \boldsymbol{j}(t) \qquad (1)$$

The depolarization current density in the case of zero conductivity is

$$j_d(t) = E \cdot [\boldsymbol{j}(t+t_o) - \boldsymbol{j}(t)] \qquad (2)$$

where $t_o$ is the poling time. If the poling time is long enough ($t_o \gg t$), then $\boldsymbol{j}(t+t_o) \to 0$, so, as follows from Eqs. (1) and (2),

$$j_p(t) = -j_d(t), \qquad (3)$$

i.e. $j_p(t)$ and $j_d(t)$ should be mirror images of each other. Therefore, the difference between $j_p(t)$ and $j_d(t)$, if observed experimentally, can be caused by either a finite conductivity ($g \neq 0$), or by incomplete poling due to a relatively short poling time $t_o$. In both cases one would observe $j_p(t) \neq j_d(t)$.

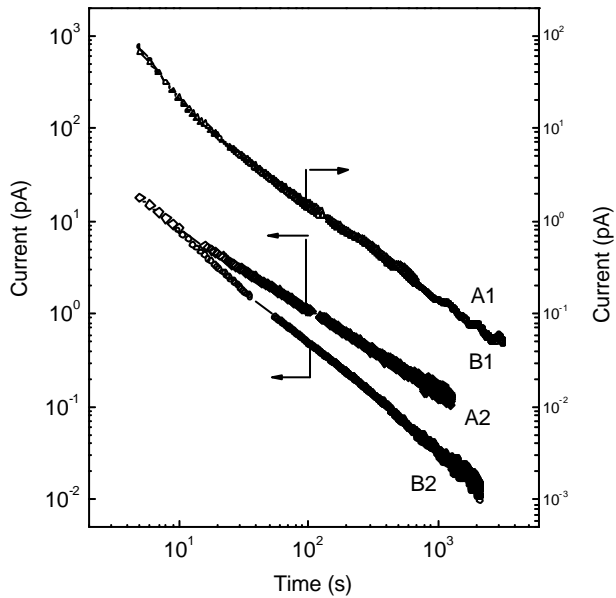

**Fig. 1** Isothermal polarization (A) and depolarization (B) currents: A1 - 50 °C, 3000 s; B1 - 50 °C, 3000 s; A2 - 25 °C, 1000 s; B2 - 25 °C, 2000 s.

To distinguish between influence of the two factors, one should analyze the shape of the polarization current at long times $t$. In the case of absence of the conductivity and long poling times, $j_p(t)$ and $j_d(t)$ coincide with both currents going to zero. If the sample has a definite conductivity, the through constant current will be observed at the $j_p(t)$ curve, while the $j_d(t)$ curve will go to zero. It follows from the above mentioned that by comparing isothermal polarization and depolarization currents one can get information on the poling time required to obtain the sufficient polarization and on the value of the conductivity.

Experimentally measured polarization and depolarization currents are presented in Fig. 1 and 2. At room temperature and long poling times (Fig. 1, curves A1 and B1), polarization and depolarization currents coincide indicating that the conductivity is very low and can be neglected at these conditions. If the poling time is not long enough, a deviation of $I_d$ from $I_p$ is observed, as in the case shown in Fig. 1, curves A2 and B2, being in accordance with Eqs. (1) and (2).

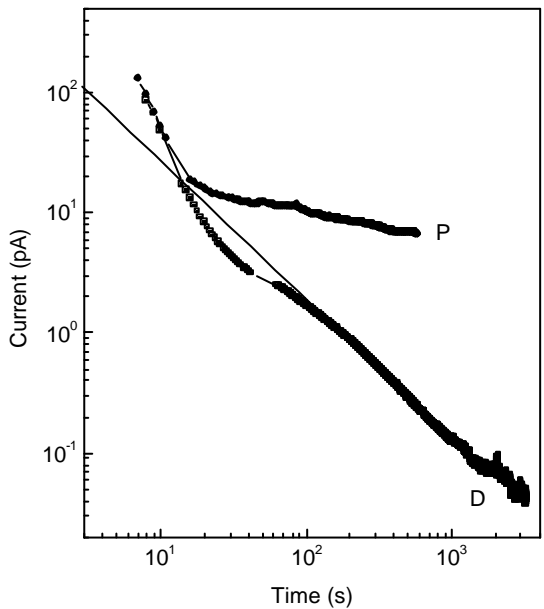

Fig.2. Isothermal polarization (P) at 90 °C for 600 s and depolarization (D) at 90 °C. Sample PS + 2% DR1.

It has been found that the conductivity can be neglected provided the poling temperature $T_p$ is lower than the glass transition temperature $T_g$=90 °C. Even at 80 °C there was no tendency for the polarization current to reach the saturation. However, if $T_p>T_g$, as one can see from Fig. 2, the polarization current is much higher than the depolarization one and it tends to become constant at long poling times (purely conductive current).

It is known that in the Debye case of all dipoles having the same relaxation time t, the decay of polarization and depolarization currents should correspond to the exponential function $I(t)=I_o exp(-t/t)$. The curves shown in Fig. 1 and 2 consist of at least two parts and match rather the power law $I(t) \sim t^{-n}$ with $n \gg 2$ at $t$ from 0 to 10 s followed by a protracted region where $n \approx 0.9$-1.0. This feature is usually considered as an indication of a broad distribution of relaxation times. Straightness of the $I_d(t)$ curve in the whole range of time with $n > 1$ can indicate incomplete poling.

Isochronal values of polarization and depolarization currents depend on temperature. This dependence is more sound at small times ($I_p(80 °C) = 60$ pA, $I_p(25 °C) = 8$ pA at $t = 10$ s), but then the effect of temperature becomes negligible ($I_p(80 °C) \approx I_p(25 °C) = 0.1$ pA at $t = 1000$ s).

External injection of charge carriers is not likely to effect the conductivity in our case, since the current - voltage characteristic (not shown here) is linear at $T>T_g$ and isothermal polarization and depolarization currents are symmetrical below $T_g$ as shown in Fig. 1. Therefore, the space charge is formed locally due to spatial separation of intrinsic carriers. Separated charges are most probably trapped between polymer chains forming virtual induced dipoles. Near $T_g$, the trapped charges are shaken out due to increased mobility of the fragments of the polymer chains but they do not move far under action of the poling field, but rather recombine, as if the local field is higher than the external one. It is clear that the response function of doped polystyrene is not the exponent indicating that processes of polarization build-up and its relaxation even in the isothermal case are more complicated than those for which a simplified model, such as the Debye one, can be constructed.

It is interesting to find out whether the material under study behaves electrically as a linear system, or it shows not only non-linear optical properties, but also the electrical ones.

### 3.2. Superposition principle

In the case of the electric field $E$ applied to a dielectric, the superposition principle establishes that if the response function $\boldsymbol{j}$ (t) for a unit step variation of the electric field, as well as the time dependence of the electric field are known, then it is possible to calculate the current $I(t)$ at any moment according to the following equation

$$I(t) = A\int \boldsymbol{j} \ (t-\boldsymbol{q})\frac{dE}{d\boldsymbol{q}}d\boldsymbol{q} \qquad (4)$$

where A is the surface area of the sample. Suppose a step voltage is applied producing a constant electric field $E_o$ during a period of time $t_o$ from $t = -t_o$ to $t = 0$ and after that the sample is short circuited. In that case one can write

$$\frac{dE}{dt} = E_o\boldsymbol{d}(t + t_o\ ) - E_o\boldsymbol{d}(t) \qquad (5)$$

where $\boldsymbol{d}(t)$ is Dirac's delta function. Therefore, the Eq. (4) for the transient current can be rewritten for $t > 0$ as

$$I(t) = AE_o[\int \boldsymbol{j} \ (t-\boldsymbol{q})\boldsymbol{d}(\boldsymbol{q} + t_o)d\boldsymbol{q} - \int \boldsymbol{j} \ (t-\boldsymbol{q})\boldsymbol{d}(\boldsymbol{q})d\boldsymbol{q} \ ] \qquad (6)$$

After integration and considering the depolarization current as a positive one we obtain

$$I(t) = AE_o[\boldsymbol{j} \ (t) - \boldsymbol{j} \ (t + t_o)] \qquad (7)$$

It is clear from Eq. (7) that if the poling time $t_o$ is very long ($t_o \circledR ¥$), then $\boldsymbol{j} \ (t + t_o) \rightarrow 0$ and $I_¥(t)=AE_o\boldsymbol{j} \ (t)$. Therefore, if the superposition principle is applicable, it is possible to calculate the depolarization current corresponded to any poling time $t_o$ from only one experimental curve $I_¥(t)$. Thus, the opposite is also accurate, i.e. if the experimentally measured and calculated isothermal depolarization currents match with each other, then the superposition principle is valid and the system is linear.

Corresponding calculations were performed for samples of PS+1.5%DR1 at different temperatures below $T_g$ and results of the calculations were compared with experimental $I_d(t)$ curves. As an example, the curves obtained at 75 $^o$C and 90 $^o$C are shown in Fig. 3 and 4. To obtain the experimental curves we applied the poling voltage of 100 V for a definite time $t_o$ in the range from 4 s to about 1000-2000 s and then measured the depolarization current. The current corresponded to the longest poling time was taken as $I_¥(t)=E_o\boldsymbol{j} \ (t)$ for further calculations according to Eq. (7).It is clear from Fig. 3 and 4 that even for a very short poling time of 4 s, good agreement is observed between experimental and calculated data, proving validity of the superposition principle for a given system. This result is neither obvious nor trivial, because the polymer under study possesses nonlinear optical properties. At the same time, as our results have proven, the doped PS can be considered electrically as a linear system.

From the data shown in Fig. 3 and 4 one can conclude also that the response function $\boldsymbol{j}$ (t) deviates considerably from the often postulated power law of $\boldsymbol{j}$ (t) ~ $t^n$, but at low and medium poling times the curve $I_d(t)$ is very close to the straight line in logarithmic coordinates. In general, the depolarization current curve consists of three parts. The current at the first stage (from $t = 0$ till $t \cong 10$ s) does not depend much on the poling time indicating that it originates from depolarization of dipoles having short relaxation times. In the power law approximation, $n \approx 2$ is usually observed at this stage. At the second stage (from 10 s to about 100-200 s) the curves obtained at different poling times go apart with $n$ ranging from 2 at very short poling times to 1 for very long poling times. Therefore, at this stage the shorter poling time, the smaller the power coefficient $n$. At the third stage, absolute values of the currents corresponded to different poling times are very different, but the coefficient $n$ is almost the same for all curves which are now parallel to each other in logarithmic coordinates with $n \approx 2$, similarly to that at the first stage.

It follows from Eq.(7) that isochronal values of the depolarization current must be proportional to the poling field. This feature of the superposition principle was examined by comparing experimental depolarization curves obtained after poling for the same time ($t_o = 20$ s in our experiments), but at different poling voltages from 30 V to 300 V. The results have shown that the proportionality of the isochronal currents to the poling field is actually observed additionally proving the validity of the superposition principle.

Isothermal depolarization current $I_d(t) = A\dfrac{dP}{dt}$ reflects dynamics of the isothermal relaxation of the residual polarization $P$. Since the depolarization experiments are performed under short circuit conditions ($E=0$), the conductivity phenomenon does not effect the measurements. Constant voltage was applied for a rather long time and then the sample was short circuited at a definite moment and the depolarization current was recorded.. The results are shown in Fig. 5.

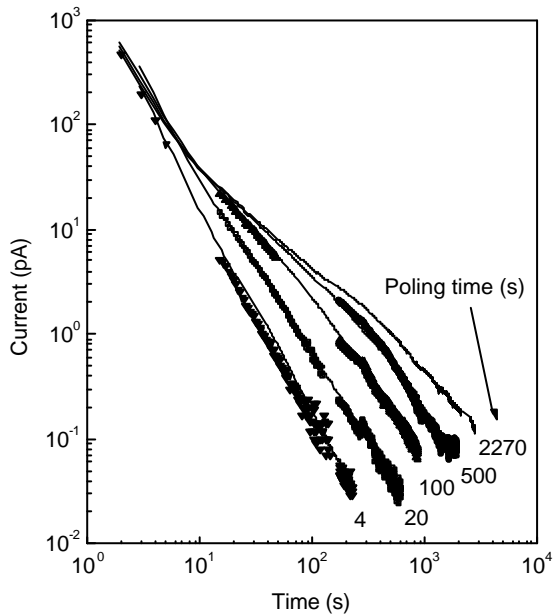

Fig.3. Isothermal depolarization currents at 75 °C in isothermally poled PS+1.5% DR1 samples with different poling times Solid lines - experiment, points - calculations.

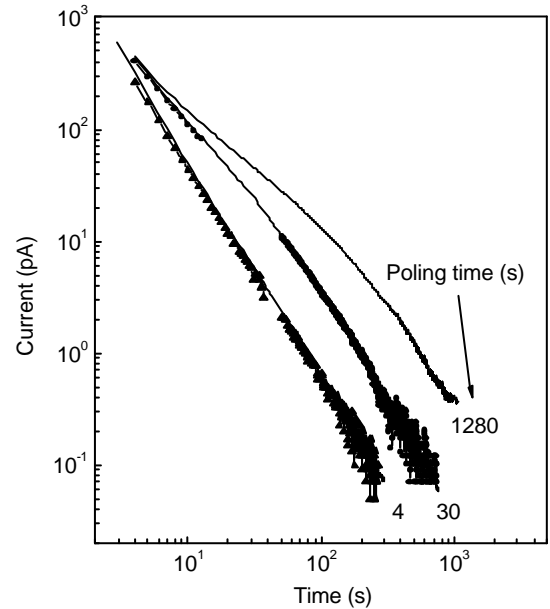

Fig.4. Isothermal depolarization currents at 90 °C in isothermally poled PS+1.5% DR1 with different poling times. Solid lines - experiment, points - calculations.

It is seen from Fig. 5 that curves for 30 °C and 50 °C coincide, while the curve obtained at 70 °C is parallel to that of 30 °C and 50 °C indicating qualitatively that the relaxation function remained unchanged, but the level of the residual polarization and consequently, the rate of its decay (the depolarization current) increased with temperature in the range from 50 °C to 70 °C. It is possible that the dipoles are easily oriented at higher temperatures due to decreasing of the relaxation time. It means also that the saturated value of the polarization has not been reached during poling, because the equilibrium polarization, according to the Langevin-Debye equation, should decrease inversely proportional to temperature. However, in this fact happens at $T>T_g$. Due to intensified motion of the polymer chains, the thermal disorienting during poling prevails over the orienting effect of the poling field. As the result, the final value of the residual polarization at the end of poling decreases. Moreover, a large number of the dipoles are disoriented during the short period of time after the short circuiting, producing the first part of the depolarization curve characterized by high values of the current and by its fast decreasing (curves for 90 °C and 110 °C in Fig.5).

PS does not show considerable conductivity at temperatures below $T_g$. On the other hand, at $T > T_g$, the current measured under a constant applied voltage increases showing existence of some conductivity. The origin of this conductivity is not definitely established, but since it is thermally activated, one can suggest that the conductivity current almost certainly originates from internal emission of intrinsic charge carriers, rather than from the external injection. This suggestion is supported by linearity of the current - voltage characteristic at $T > T_g$.

It has been found that the temperature dependence of conductivity in doped PS is dissimilar for different content of DR1, so as the higher concentration of the dopant corresponds to the higher conductivity. It is possible that together with the DR1 molecules some other species are added to PS matrix including those producing charge carriers at high temperature. As for the activation energy of conductivity, it has the same order of 2.0 - 2.3 eV.

Conductivity of the samples at $T < T_g$ can be neglected because the saturated steady current was not observed even after application of the poling voltage for a long time. Moreover, polarization and depolarization currents in well poled samples are coincide as seen in Fig. 1.

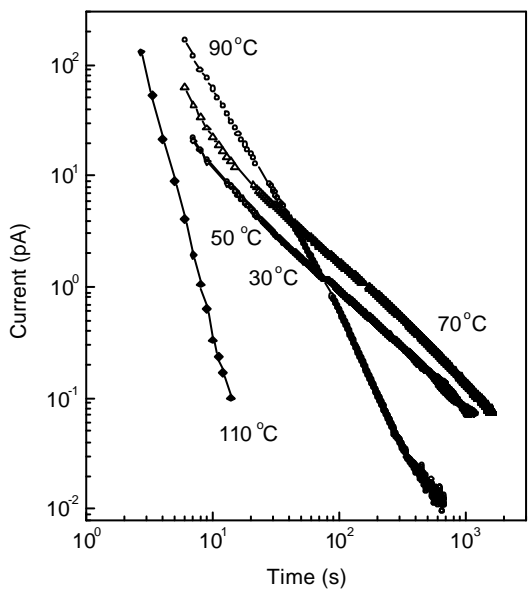

Fig.5. Isothermal depolarization of PS+1.5% DR1 after poling at 150 V at different temperatures.

## 4. Conclusion

It has been shown that the electrical relaxation of the poled order in PS doped with DR1 chromophore molecules can be successfully studied by measuring isothermal polarization and depolarization currents. It has been proved that the superposition principle is applicable to the polymer system under study, so as the doped PS can be considered electrically as a linear system with a negligible electric conductivity at temperatures lower than the glass transition temperature.

It has been found that the depolarization current curve of doped polystyrene consists of three parts with the response function being essentially non-exponential one, but rather characterized by a broad distribution of the relaxation times. It means that the behavior of even such a simple system as PS doped with DR1 is much more complicated than that predicted by known theoretical models.

By comparing isothermal polarization and depolarization currents one can get practical information on the preferable poling conditions, such as the poling time and temperature required for obtaining the adequate and stable residual polarization.

## References

1. C. C. Chang,, C. P. Chen, C .C. Chou *J. Macromol. Sci.-Pol. R.* 2005, **C35**, 125.
2. F. Kajzar, K. S. Lee, A. K. Y. Jen *Adv. Polym. Sci.* 2003, **161**, 1.
3. L. Dalton *Adv. Polym. Sci.* 2002, **158**, 1.
4. N. Bloembergen *J. Nonlinear Opt. Phys.* 1996, **5**, 1.
5. G. M. Sessler (ed.) *Electrets*, Vol. 1, Third Edition, Laplacian Press, Morgan Hill, 1999.
6. E. Cecchetto, D. Moroni, B. Jerome *J. Phys.-Condens. Mat.* 2005, **17**, 2825.
7. R. D. Dureiko, D. E. Schuele, K. D. Singer *J. Opt. Soc. Am. B.* 1998, **15**, 338.
8. A. Dhinojwala, G. K. Wong, J. M. Torkelson *J. Chem. Phys.* 1994, **100**, 6046.
9. V. Sangawar, C. S. Adgaonkar *J Polym. Mater.* 1996, **13**, 207.
10. V. Sangawar, C. S. Adgaonkar *Indian J. Pure Appl. Phys.* 1996, **34**, 101.